# The Ancient Astronomy of Easter Island: Aldebaran and the Pleiades

Sergei Rjabchikov[1]

[1]The Sergei Rjabchikov Foundation - Research Centre for Studies of Ancient Civilisations and Cultures, Krasnodar, Russia, e-mail: srjabchikov@hotmail.com

**Abstract**

There is a good cause to assert that the Easter Islanders constantly watched Aldebaran and the Pleiades in the past. The Russian scholar Irina K. Fedorova and the American scholar Georgia Lee were the first who contributed significantly to the archaeoastronomical studies of these celestial bodies. I have decoded the Mataveri calendar completely. The ceremonial platform Hekii 2 was oriented on Aldebaran and nearby stars. The disappearance of the Pleiades during the dawn period in the north at the end of August could be an important mark before the arrival of sooty terns and before the elections of the next bird-man. The calculated dates of several solar eclipses have been used for composing the chronology of Easter Island from 1771 till 1867 A.D.

**Keywords**: archaeoastronomy, writing, folklore, rock art, Rapanui, Rapa Nui, Easter Island, Polynesia

## Introduction

The civilisation of Easter Island is famous due to their numerous ceremonial platforms oriented on the sun (Mulloy 1961, 1973, 1975; Liller 1991). One can therefore presume that some folklore sources as well as *rongorongo* inscriptions retained documents of ancient priest-astronomers.

### A Farther Remark about a Rapanui Rock Calendar: A Key to the Mataveri Observatory

On a boulder at Mataveri (a crucial area of bird-man rites) some lines were incised; most of them were the directions of the setting sun according to Liller (1989). I have calculated the corresponding days for the year 1775 A.D. (Rjabchikov 2014a: 5, table 2; 2015: 2, table 1; 2016a: 1, table 1). Here and everywhere else, I use the computer program RedShift Multimedia Astronomy (Maris Multimedia, San Rafael, USA) to look at the heavens above Easter Island.

**Table 1.** The Dates Calculated (with the interpretation for October 1 and 3 as well as November 14):

**June 22** (the azimuth of the sun = 296.2°): one day after the winter solstice;
July 21 (292.5°): the star Capella (α Aurigae) before dawn;
August 11 (286.7°): the star Pollux (β Geminorum) before dawn;
September 2 or 3 (277.9°): the star β Centauri [*Nga Vaka*] before dawn;
**September 21** (270.1°): the day before the vernal equinox (the key moment of the bird-man feast);
September 24 (268.7°): the new moon;
September 27 (267.4°); the fourth night: the measure of the visible dimensions of the moon; **the waxing crescent was well seen in the sky; one night before the beginning of the first *Kokore* lunar series**;
October 1 (265.9°); the eighth night: the measure of the visible dimensions of the moon; **the first quarter of the moon**;
October 3 (264.7°): **the gibbous moon (the 10th night); the end of the first *Kokore* lunar series**;
October 22 (256.8°): near the new moon;
November 8 (250.7°): the star Spica (α Virginis) before dawn;
November 12 (249.3°); Venus as the Morning Star before dawn;
November 14 (248.7°); **one night before the last quarter of the moon**;
November 23 (246.3°): the new moon;
**December 20** (the azimuth of Aldebaran = 339.1°): the star Aldebaran (α Tauri) at night;
**December 21** (the azimuth of Aldebaran = 322.1°; the azimuth of Canopus = 177.5°): the stars Aldebaran (α Tauri) and Canopus (α Carinae) on the same night (Rjabchikov 2013a: 7); the day of the summer solstice; **one night before the new moon**.

Aldebaran is the bright red star and due to its colour this heavenly body could have the sacramental value. Canopus is the second brightest star after Sirius (α Canis Majoris). The observations of both stars together just before the day of the summer solstice were remarkable.

Since in the old days on Rapanui the New Year began with the new moon of the month *Maro* or *Maru(a)* 'June chiefly' (Rjabchikov 1993a: 133-134), the first morning appearance of Aldebaran prior to the new moon of June was a peculiar mark before the day of the winter solstice.

The big royal astronomical complex at the western area of Rapanui could include the solar observatories at the sites Orongo (Ferdon 1961; 1988) and Mataveri (Liller 1989) as well as at the ceremonial platforms Ahu Huri a Urenga and both Ahu Vinapu and Ahu Tahiri. On the last two platforms located closely near each other 15 statues stood, hence one can suggest that all they represented the phases of the waxing moon from the new moon to the full moon (*Vina Pua = Hina Pua*). So, the Vinapu and Tahiri (*tahiri* 'the elevation') platforms oriented on the sun (Mulloy 1961) were the distinctive lunar calendar device, too. The Ahu Huri a Urenga with the single statue oriented on the sunrise at winter solstice (Mulloy 1973) could be a mark of the transition of short days of winter to long days of summer. The name of this place contains the terminology of the darkness (*huri*, *uri*) and fertility (*ure-nga*; cf. also Maori *uri* 'offspring, issue') associated with the ghost *Hauriuri* (The Blackness). In this way it is necessary to understand the cultural context of the observations of the Pleiades and Aldebaran in the past.

It is of interest that another boulder in the environs of the Mataveri has several cupules that give the precise map of several stars of the Sagittarius constellation (Hockey and Hoffman 2000).

### The star Aldebaran in the Rapanui Script

Fedorova (1982) read at first glyphs **7 70** *(H)etuu Pu* 'Aldebaran.' Unfortunately, later on she gave this correct interpretation up (Fedorova 2001).

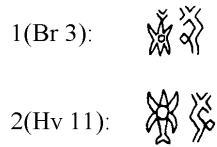

Figure 1.

The stellar name **7 70** *Tuu Pu* 'Aldebaran' is put down in the records of the Aruku-Kurenga (B) tablet and Great Santiago (H) tablet, see figure 1. The latter fragment is shown between some other signs, see figure 2. In the light of my last research, the text reads as follows.

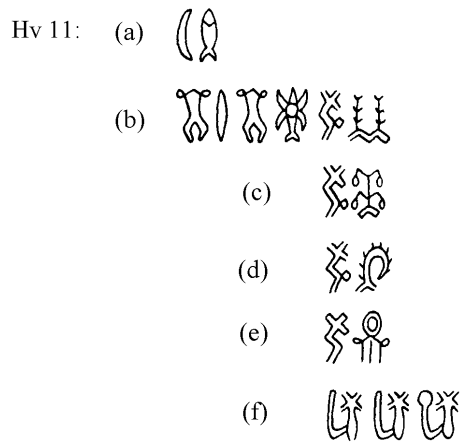

Figure 2.

Hv 11: (a) **3 12** (b) **75 30 75 7-70 24-24** (c) **70 25var** (d) **70 14** (e) **70 39 26** (f) **4-45 4-45 4-45** *Marama Ika: ko ana, ko Tuu Pu Ariari, (Tuu) Pu Hua, (Tuu) Pu Haua, (Tuu) Pu Raa maa. Tupu, tupu, tupu.* The moons of the Fish (as in the first part of the Manari calendar text): The star Aldebaran shone (on the nights) *Ariari* (= *Ari*), *Hua*, *Haua* (= *Atua*), (and) the bright *Ra*[*kau*]. The growth, the growth, the growth.

In the text there are names of four nights (moons) of the local lunar calendar. These names are compared with their related names in the famous calendar on the tablet Mamari (C), see table 2.

**Table 2.** The Comparison of Some Signs from the Two Boards

| The Mamari tablet (Ca 6-7) | 12 + 3 | 24-24 | 25a | 14 | 106 = 139 8var |
|---|---|---|---|---|---|
| The Readings | *Ika + Marama* | *Ariari* [*Ari*] | *Hua* [*Hua*] | *Haua* [*Atua*] | *Raa Kau* [*Rakau*] |
| The Great Santiago tablet (Hv 11) | 3 12 | 24-24 | 25var | 14 | 39 26 |
| The Readings | *Marama Ika* | *Ariari* [*Ari*] | *Hua* [*Hua*] | *Haua* [*Atua*] | *Raa* [*Rakau*] *Maa* (Bright) |

The key readings of glyphs **24-24** *Ariari* and **14** *Haua*, the 4th and 13th nights (moons), are presented in Rjabchikov 1989: 123-124, figure 3, fragments 1 and 3. The names *Ariari* and *Haua* are associated with the names *Ari* and *Atua* in the calendar list by Métraux (1940: 50). The first case is obvious: the form *Ariari* is the complete reduplication of the form *Ari*. In the second case the moon was an incarnation of the moon goddess *Hina* known also as *Haua*, the wife of the solar god *Tiki-Makemake*, and the words *atua*, *etua* meant 'god; goddess.' The scribes therefore substituted the word *Atua* (the Goddess) for the divine name *Haua*. Glyph **24** represents the garland and reads *ai* as well as *ari* because of the alternation of the sounds *r*/-  (cf. Rapanui *hei* 'garland,' Rarotongan *'ei* 'ditto' and Kapingamarangi *hai*, *hei* 'to put on;' the variations of the sounds *h*/-, *e*/*a* are very frequent as well). Glyph **14** represents the headgear (cf. Rapanui *hau* 'hat'). The scribes chose such a shape of these signs that resembled the crescent.

One can read this inscription on the Aruku-Kurenga tablet, see figure 3.

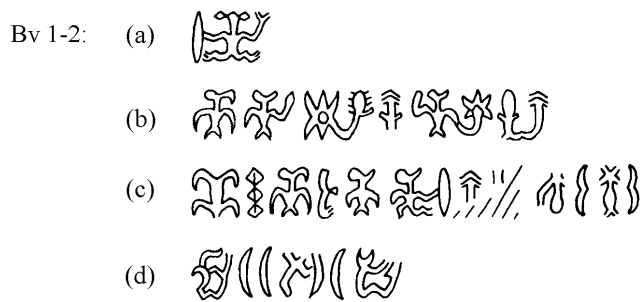

Figure 3.

(a) **30 69** *Ana Moko.* (It was) the house of the Lizard [at the sacred village of Orongo].
(b) **44-44 7-25 9 33 44 7-25 33** *Tahataha Tuu Hu niva ua. Taha Tuu Hu ua.* The star Aldebaran (appeared first in the rainy season) turned. The star Aldebaran (appeared first in the rainy season) turned.
(c) **80 17 44 43 44 6 30 33 9 45-50-45-50** *Ui te tamata Hanau Niva Puhipuhi.* The man (*tamata* = *tangata*, cf. also Maori *tama* 'man') *Puhipuhi* from the Miru tribe watched it.

(d) **3 50-15 3 3 70 140 6 102** *Hina Ira marama, marama Pua* "the fill moon," *a Ure*. (They are the nights) *Ina Ira*, the full moon *Pua* (= *Hina-Pu*, *Vina-Pu*), *Ma-ure*.

In accordance with this text, the priest *Puhipuhi* served inside a stone house (*ana*) of the religious centre of Orongo. That house was dedicated to the god *Hiro*; per the local beliefs, the Lizard was an incarnation of this deity (Barthel 1978: 251). The duties of *Puhiputi* were the predictions of the whether and of eclipses as well as writing calendar records. *Tuu hu* 'Aldebaran' was a variant of the stellar name (Rjabchikov 1993b: 6, table 3). *Puhipuhi* was a known person, and his name could be retained in the people's memory. *Kopuhi Angataki* was a famous wizard *ivi atua* (Brown 1996: 125). The particle of personal names, *ko*, is added to his name. So, the well-known astronomer *Puhipuhi* was called *Ko Puhi anga taki* = '*Puhi* (Wind) who predicts (*taku*) and counts (*tataku*).' One can mention Maori *taki* 'to repeat; to multiply' also. I surmise that his another name was *Rangi Taki* (Felbermayer 1971: 29-32).

Having compared line (d) with line Ca 7 of the calendar text inscribed on the Mamari tablet, one can conclude that natives omitted one day (night) of 30 days of the month not only in the end of the month, but also in the centre, choosing, for instance, the night (moon) *Maure*. The while was diminished one time during two months to achieve the average length of the lunar month approximately 29.5 days.

Popova (2015) has made a great contribution to the studies of the archaic Rapanui beliefs about the star Aldebaran.

## The Pleiades in the Rapanui Script

Lee (1992: 80) reasons that in some cases the turtle represents the stellar cluster of the Pleiades (M45; NGC 1432). In her opinion, the ancient towers *tupa* were exploited to watch these stars (Lee 2004: 36). On the basis of my own research I have deduced that this hypothesis is correct; on a Rapanui panel are rendered the drawing of a turtle, seven cupules as well as one stylised glyph **49** *(ariki) mau* (Rjabchikov 2001: 218).

Consider a record on the Tahua tablet (A), see figure 4a.

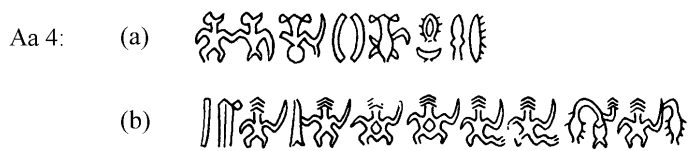

Figure 4a.

In segment (a) the rising (glyphs **6-44** *hata*, cf. Maori *whata* 'to elevate') of the Pleiades (glyphs **68** *Honu*, **68** *Honu*) before dawn (glyph **57** *tara*, cf. Maori *tara* 'rays of the sun, shafts of light, appearing before sunrise') on the 13th night/day (glyphs **3 14** *marama Haua = Atua*) of the lunar month is described. In segment (b) seven combinations of glyphs **6-33** *hau* (king) are inscribed (although the inverted reading is also conceivable: **33-6** *uha* 'woman'). Here glyph **12** *ika* may denote a cluster (Maori *ika* 'ditto'), and two glyphs **25** *hua* (fruit) may be a hint at one Maori name of the Pleiades, *Hoko-kumara* (Best 1922: 33), cf. Rapanui *oko* 'ripe.' Alternatively, glyphs **25** *hu(a)* may denote the star Aldebaran.

The natives of different Polynesian islands saw in the sky this cluster consisting of seven and even of six stars. The next explanation is available: cf. Rapanuio *honu*, Marquesan *hono* 'turtle,' and Rapanui *ono* 'six.' On Easter Island six boulders were a model or a map of the Pleiades (Métraux 1940: 53). A Mangaian myth tells that god *Tane* broke one brilliant star with the assistance of Sirius and Aldebaran, and six shining bits formed the Pleiades (Gill 1876: 43).

In conformity with the mythology of the inhabitants of Pukapuka (the Cook Islands), the god *Matariki* was son of *Tamaei* who came there from Tonga (Gill 1912: 122-123). I presume that this *Tamaei* (*Tama 'ei* 'The Forefather decorated with garlands') still was the potent sun god *Tama* (Father, Ancestor) known later as the god *Tangaroa* of the sea, sailors and fishermen (Rjabchikov 2016b, 2016c).

According to the Hawaiian mythology, the name *Makari'i* [*Matariki*] reads *Maka ari'i* [*Mata ariki*] 'Eyes of the Chief' (Beckwith 1970: 367). In the Maori beliefs the Pleiades were seven chiefs (Tregear 1891: 226). Moreover, the Maori data witness that the Pleiades are *Mata-ariki* (*mata*: face; *ariki:* lord) (White 1887: 5).

Consider now a record on the Santiago (I) staff that once belonged to king *Nga Ara*, see figure 4b.

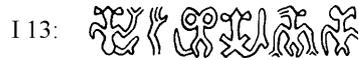

Figure 4b.

I 13: **68 (102) 15 4 60 49c (102) 6-44 …** *Honu: roa atua Mata ariki (mau), hata…* (They were) the Pleiades: the deity (deities) of the Pleiades lifted himself (themselves), (it was) the elevation…

Notice that glyphs **102** *ure* were inserted here as determinatives (fertility, abundance). This text has been partially decoded earlier (Rjabchikov 2001: 217-218, figure 2).

**The Linguistic Background**

Glyph **49** reads *mau* and *ariki mau*, so I point this reading as *(ariki) mau*. But in some cases it reads *ariki*, and I point it as *ariki (mau)*. This glyph represents king or chief. This person wore a hat indicated with feathers, and in many cases the headdress is put on the head of the sign. Besides, in several cases glyph **41** (different variants) is represented instead of his legs. This latter glyph reads *re* meaning 'winner.' Even if glyph **49** contains glyph **41**, the whole sign reads *(ariki) mau*, or *ariki (mau)*, or *ariki mau* (Rjabchikov 1998). Typologically, in the Maya script some glyphs include of parts of other glyphs.

In compliance with Métraux (1940: 231), the wooden pendants *rei-miro* were in the midst of the royal insignia. Consider the brief record on the London *rei-miro* (J), see figure 5.

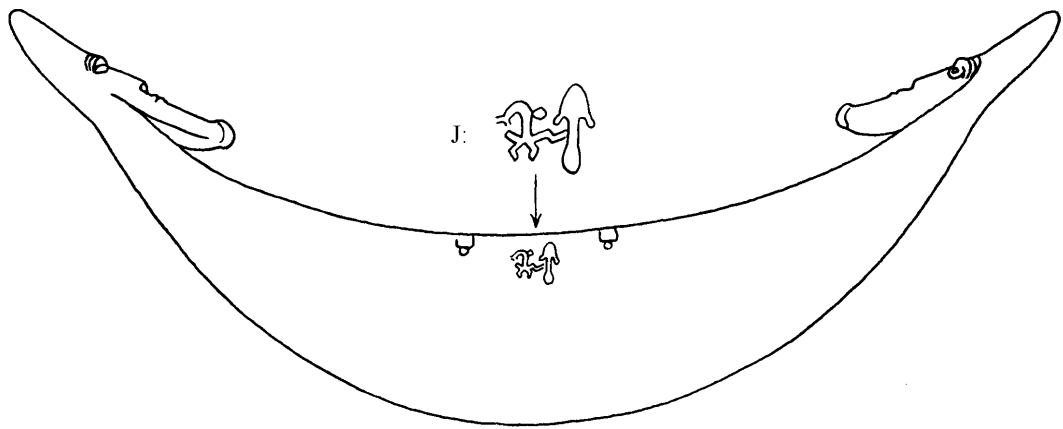

Figure 5.

J: **49a 22** *Ariki mau ao*. The king has authority.

No name was inscribed, so this object was a symbol of high rank and a tribal sign of the fertility in the possession of several kings.

Consider the record on the Small Washington tablet (R), see figure 6.

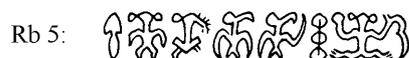

Figure 6.

Rb 5: **22 6 49a 44-44 17 69-50** *Ao a ariki mau tahataha te Makoi*. King *Makoi* [*Kai-Makoi* the First] died (= the authority of king *Makoi* disappeared literally).

Rapanui *taha* means 'to set (of sun).'
Consider a record on the Santiago staff, see figure 7.

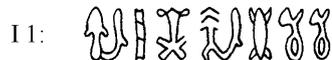

Figure 7.

I 1: **22 (102) 4 49c 59-33 (102) 28 6-39-6-39** *Ao atua, (ariki) mau, kaua Nga Araara.*
*Nga Araara* (= *Nga Ara*) who is the lord, the king (and) the progenitor has authority.

In all three records both terms, *ariki mau* 'king' and *ao* 'authority,' are rendered. Butinov and Knorozov (1957: 10,13, 16 , table 2, fragment 10; table 5, fragment 10; table 8, fragment 2) have read glyph **49a** *ariki* (king; chief) on the basis of the "readings" of the native *Metoro*. In my studies of the *rongorongo* I have ignored his information as unreliable in the majority of cases.

### The Continuation of Our Studies of the Information about the Pleiades

Consider another record on the Aruku-Kurenga tablet, see figure 8.

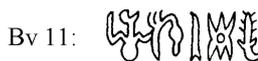

Figure 8.

Bv 11: **68 15-25 5 7-25** *Honu (Hono, Ono) ro(h)u atua Tuu Hu*. The Pleiades bore the deity Aldebaran.

The main idea of this observation was such: the Pleiades as a share of the Taurus constellation always appear in the sky before Aldebaran. Old Rapanui *ro(h)u*, *ra(h)u* 'to create' (glyphs **15-25, 27**) correspond to Samoan *malaulau* 'to grow vigorously' and Maori *whakarau* 'to multiply.'
Now one can investigate groups of glyphs **68** written together with subsidiary glyphs on the London (K), Mamari, Great Washington (S), Small Santiago (G) and Keiti (E) tablets, see figure 9.

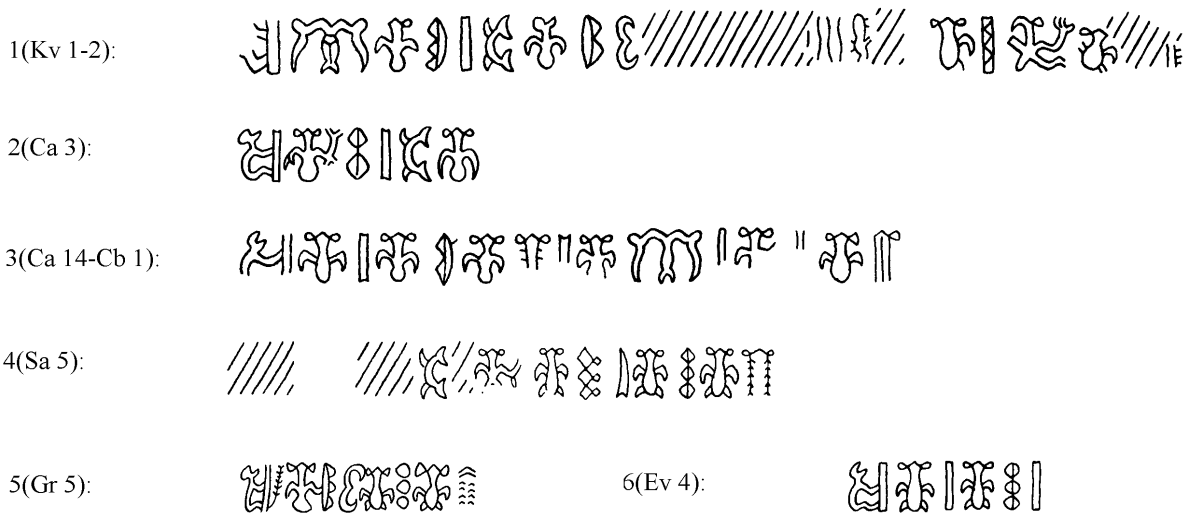

Figure 9.

1 (Kv 1-2): **6-4 44-12 68 18 4 8 68 18** (a damaged segment) **4**? **4**? **25**? or **73**? (a damaged segment) **68 4 49a 68** (a damaged segment) **46**? or **4**?
2 (Ca 3): **6-4 68 15 17 4 8 68**
3 (Ca 14-Cb 1): **6-4 68 4 68 18 68 56 4 68 44-12 68 15 4 68 26**
4 (Sa 5): (a damaged segment) **8** (a damaged segment) **68 68 17 5 68 17 68 24-24**
5 (Gr 5): **6-4-24 68 18 23 68 17 68 24-24**
6 (Ev 4): **6-4 68 4 68 17 4**

In five cases the texts begin with glyphs **6-4** *a atua* 'deity,' and in one case the expression **6-4-24** *a atua ai* 'deity-place' is taken down. These glyphs as well as glyphs **4-32** *atua ua* and **6-4-32** *a atua ua* 'deity-dwelling' denote names of deities, corresponding statues (*moai* < *mo ai* 'for the place') and other sacred places (Rjabchikov 1988: 316-317, figure 3, fragment 9; 2001: 216-219, figures 1 and 4; 2009a, figures 2 to 4, 6 to 10, and 39; 2014b: 172-173, figure 9, fragment 6). Side Ca = Cr and side Cb = Cv (Rjabchikov 1995: 51-52).

A part of the inscription in fragment 1 has been investigated earlier (Rjabchikov 2000: 68, figure 5). There I disclosed two symbols (glyphs **68**) of the Pleiades. I suppose that the god *Atua-Matua* (glyphs **4 8**) was associated with the Maori god *Tane-matua* '*Tane*-parent' (Rjabchikov 2011a: 8-10) and with more archaic god *Tama* (*Tama loa* and so on). *Tane* as the personification of the sun who had broken the initial radiant star into six pieces was a paramount personage. We must realise now the names of six or even seven stars of the Pleiades. Best (1922: 32-33) reports the following data of the Maori: *Matariki* 'the Pleiades,' *Te Huihui o Matariki* 'ditto,' *Ao-kai* 'ditto,' *Hoko-kumara* 'ditto' as well as the names of six stars of the Pleiades: *Tupua-nuku*, *Tupua-rangi*, *Waiti*, *Waita*, *Waipuna-a-rangi*, *Ururangi*. One can recognise these components in the last six names: *tupu*: Maori *tupu* 'to grow; to increase;' *nuku*: Maori *nuku* 'distance,' *rangi*: Maori *rangi* 'sky;' *wai*: Maori *wai* 'water;' *puna*: Maori *puna* 'spring of water;' *ti*: Maori *tia* 'to stick in;' *ta*: Maori *ta* 'to strike, to beat;' *uru*: Maori *uru* 'head; chief; top.'

**Table 3.** Different names of six stars of the Pleiades taken from figure 9

| 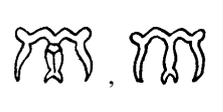 | 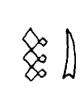 | 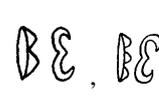 | 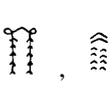 | 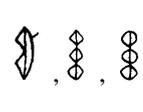 | 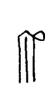 |
|---|---|---|---|---|---|
| **44-12** *ta ika* beaten, broken | **17 5** *te ti* pierced; beaten (*ati*) | **18 23** *te uru* the upper part, the head | **24-24** *ariari* visible, clear | **18** or **17** *tea* white, clean | **26** *maa* clean, bright |

The supplementary terms in the texts in figure 9 are written with glyphs **4** *atua* 'deity,' **49** *(ariki) mau* 'king,' and **56** *po* 'night.' The terms *ari*, *tea* and *maa* describe the increasing light of the sun, the moon, and stars. So, *Ariari*, *Tea* and *Maa* are the brightest ones among all the stars of the Pleiades.

A Tahitian song (Caillot 1914: 74-75) contains the Pleiades' name in the form *Matarii ma* [*Matari'i maa*] giving the word *ma* without any translation. This phrase indeed means 'The bright (= well visible) Pleiades.' Thus, the interpretation of glyph **26** *maa* in table 3 is true.

In my opinion, different celestial bodies are described in the Rapanui song "*Ka turu ki hare mori-tae pura e*" (Campbell 1999: 210). The observations occurred in December (*o Koro, o Matua e*). The Pleiades are called *Hare o Uru* (The house of *Uru*) and *Pahere*, cf. Rapanui *pahera* 'shell of turtle.'

To offer correctly the further information, it is necessary to say some words about the war that occurred on island about 1682 A.D. The reasons of that situation are understandable because of the economic factors. The timber was close to the end, and the serial production of statues stopped, and in this condition the aristocratic eastern union could not exploited the western craftsmen and other commons as in old days. In this connection, let us interpret the results of the recent archaeological excavations at the ceremonial complex Ahu Uurauranga te Mahina (Ayres, Wozniak and Ramírez 2014). I surmise that Statue UU-15 was transported at the end of the Middle period. It could be the image of the dark moon. It had comparatively small dimensions and was without well elaboration because the king of the eastern tribes demanded to cut and erect this idol as soon as possible. The statue was transported to the Urauranga te Mahina (The Lobster-the Moon) since this area was located near the border with the western tribes. That territory which was included in the next Ahu Akahanga as the single ceremonial centre played the significant role in the bird-man cult (Rjabchikov 2009b). It is clear that there were too few logs, and the statue stayed to lie near the ceremonial platform. The archaeological dating of this statue from 1650 to 1700 A.D. is essential. It is indirect evidence that the timber vanished at that time and the war between the western and eastern tribes happened. It was perhaps the last monument produced at the Rano Raraku quarry.

The addition data about that terrible war are presented in the record put down by a scribe of the Tupa-hotu tribe (Rjabchikov 2012a: 567) on the Mamari tablet, see figure 10.

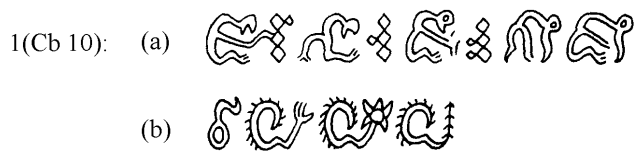

1(Cb 10): (a)

(b)

Figure 10.

1 (Cb 10): (a) **62-17 62-17 62-17 44-44 (b) 108 115 14 15 14 7 14 24** (a) *Toa Tea, Toa Tea, Toa Tea tahataha. Hiri Taka hau roa, hau Tuu, hau ari(k)i*. Many warriors of the eastern (part of the island) perished (died, disappeared etc.). *Taka* [= *Mataka roa*, i.e. the great *Mataka*] who was the great king, the king of the *Tuu* (the western tribal union), the king-(paramount) chief elevated himself.

To gain a better insight into all these celestial names of the six stars of the Pleiades, consider the following fragment of the same tablet, see figure 11.

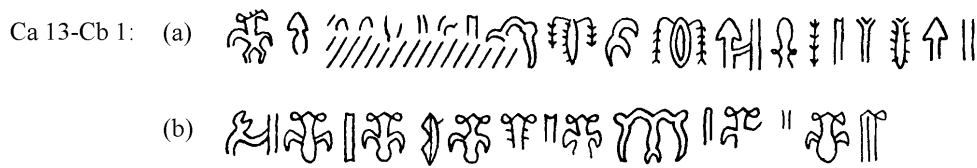

Ca 13-Cb 1: (a)

(b)

Figure 11.

Ca 13-14, Cb 1 (= Cr 13-14, Cv 1): (a) **49d 22 … 4 4-35 4 44b 24-30-24 2** (= the inverted version) **24-28-24 22 4 73 24 4 15 9 22 4** (b) **6-4 68 4 68 18 68 56 4 68 44-12 68 15 4 68 26** (a) *(Ariki) mau ao … atua Tupa, atua Tua Arina ari(k)i. Ina Arina ari(k)i ao, atua, he ai atua roa Niva ao, atua:* (b) *a atua Honu atua, Honu tea, Honu po atua, Honu ta-ika, Honu roa atua, Honu maa.* (a) The king having the authority… The lord of the Tupa-hotu tribe, the lord of the western tribes (*tua* = the back literally), (called) king *Aringa*... King *Aringa* who had the power (and) was the lord died, (then he) was in the (legendary land) Hiva (Niva literally = the West, the Sunset, Havaiki) of the power and the supremacy; (b) (it happened) near the ceremonial platform Ahu Hoonu [Ahu o Honu] (all its statues were dedicated to the stars of the Pleiades).

This text says that the king (*ariki mau ao*) Aringa (*Tupa-Aringa-Anga*) of the whole island was killed near the ceremonial platform Ahu Hoonu [Ahu o Honu] (cf. Rapanui *honu* 'turtle'), and in the local beliefs his soul travelled from that spot to the ancestral distant land called Hiva (Niva). The bay Hanga Hoonu [Hanga o Honu, Hanga Honu], related to this platform, is La Pérouse Bay.

The Juan Haoa manuscript published in the works of the Norwegian expedition (Heyerdahl and Ferdon 1965: figures 145-146) contains some interesting additional details about that time:

(1) *Koe e paoa e he oho mai ananake ko kua Mataka roa.* (2) *He oho mai ki Haga Hoonu, i ira te papaku.* (3) *Ku tuu ro ai, ku teretere ro ai: "I toona ao, iaia te ao!"* (4) *He ravaa, he tuitui, he avai ki a Mataka roa.* (5) *He oho mai ananake ki te hare Toke Takapau, i avai era tau moa era ki a Mataka roa.* (1) All the guards went together with *Mataka roa* and his companions (*kua*). (2) (They) went to Hanga Hoonu (Hanga o Hoonu), the corpse was there. (2)

(The people) came, run (saying): "(It is) his authority, he (is) the authority!" (They) seized, bound, (and) gave (that corpse) to *Mataka roa*. All they went to the house Toke Takapau (the house at Orongo where the statue Hoa-hakananaia stood)[1] and the priest (*tau* = *taua*, *taura*) gave *Mataka roa* a chicken. (The translation is of mine.)

Hence, king *Aringa* (*Tupa-Aringa-Anga*) was killed at the bay Hanga Hoonu (Hanga o Hoonu, Hanga Honu) somewhere near a ceremonial platform at this area. Only after that moment *Mataka roa*'s authority (*ao*) was proclaimed over the island. *Aringa*'s name was prohibited (cf. the mention about the unnamed bounded corpse, a "gift" for *Mataka roa*). It is of value to note that the main priest of the religious centre of Orongo blessed the new king of Easter Island.

Now it is apparent that all the records in figure 9 describe the deity (deities) (**6 4** *a atua*) of the Pleiades. Moreover, the statues on that platform at the bay Hanga o Hoonu represented the Pleiades.

According to Wallin and Martinsson-Wallin (1997: 1), the ceremonial platform Ahu Hekii 1 at this bay was called *Puapua*, *Hanga O Onu*, *Hanga-o-honu* and *Hekii*. It had seven statues. The platform Ahu Hekii 2 located alongside had four statues. I suggest that Ahu Hekii 1 (described in the Mamari record read above) was dedicated to the Pleiades. Van Tilburg (1986: 9) says that the seventh monument was in fragment, so, I think, it could be a hint at the archaic myth about the origin of the Pleiades from the broken star. In any case, the seven statues can comply with the seven stars of the Pleiades. As among the names of this complex *Puapua* was called, one can suggest that the platform Ahu Hekii 2 was devoted to the star Aldebaran (Tuu Pu) and other stars of the Taurus constellation.

Liller (1991: 275) reports about the orientation of that *ahu*: $76.0^0$ azimuth. As Hekii 1 was dated back to about 1200 to 1390 A.D. (Wallin and Martinsson-Wallin 1997: 14), I have looked at the sky those years. The azimuth of the rising Aldebaran on June 21, 1200 A.D. and June 21, 1300 A.D. was $73^0$ 51' 55" and $73^0$ 33' 20" respectively. Hence, Hekii 2 was oriented on Aldebaran and nearby stars.

Now one can examine another record about the Pleiades on the Tahua tablet, see figure 12.

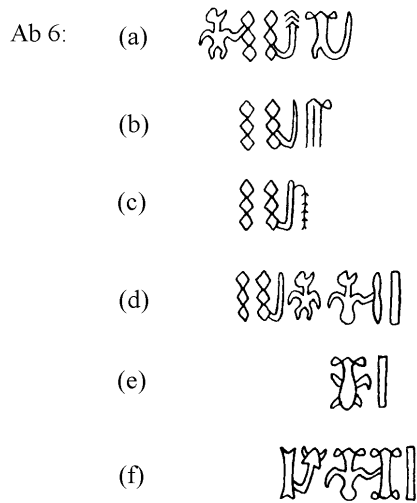

Ab 6: (a)
(b)
(c)
(d)
(e)
(f)

Figure 12.

Ab 6: (a) **6 17-17 4/33 75 5** (b) *17-17 5 26* (c) **17-17 5 24** (d) **17-17 5 6 68 65 4** (e) **68 4** (f) **4 21 68 56-56 4** (a) *Ha teatea atua ua ko Ti,* (b) *teatea atua Maa,* (c) *teatea atua Ari,* (d) *teatea atua a Honu Rangi atua,* (e) *Honu atua,* (f) *atua ko Honu po, po, atua.* This text tells of the rising of the Pleiades. Old Rapanui *teatea* 'to appear; to shine; to be visible' corresponds to Rapanui *tea* 'to appear (of stars); white, clean.' *Ha* is the verbal particle of the past tense. Here six stars of the Pleides are described, at the beginning three names are *Ti, Maa, Ari* (= *Ariari*) as in figure 9, the last three names contain the word *Honu* (Turtlel; Six = the Pleiades). The term *rangi* 'sky' was presented in one of Maori names of those stars. Besides, Hawaiian *Ka-lalani-a-Makariki* 'the Pleiades' (Pukui and Elbert 1986: 121) contains the term *lalani* (*lani*) [*rangi*] 'sky.' The names of the cluster are introduced in the text with the particles *ko* and *a*. It should be remembered that the Tahua tablet was a lesson book in the *rongorongo* school of king *Kai Makoi* the First (Rjabchikov 2012a: 566). One of the texts in the Rapanui manuscripts begins with the words: *Ko Make-make. A Makemake.* 'Makemake. Makemake' (Fedorova 1965: 397).

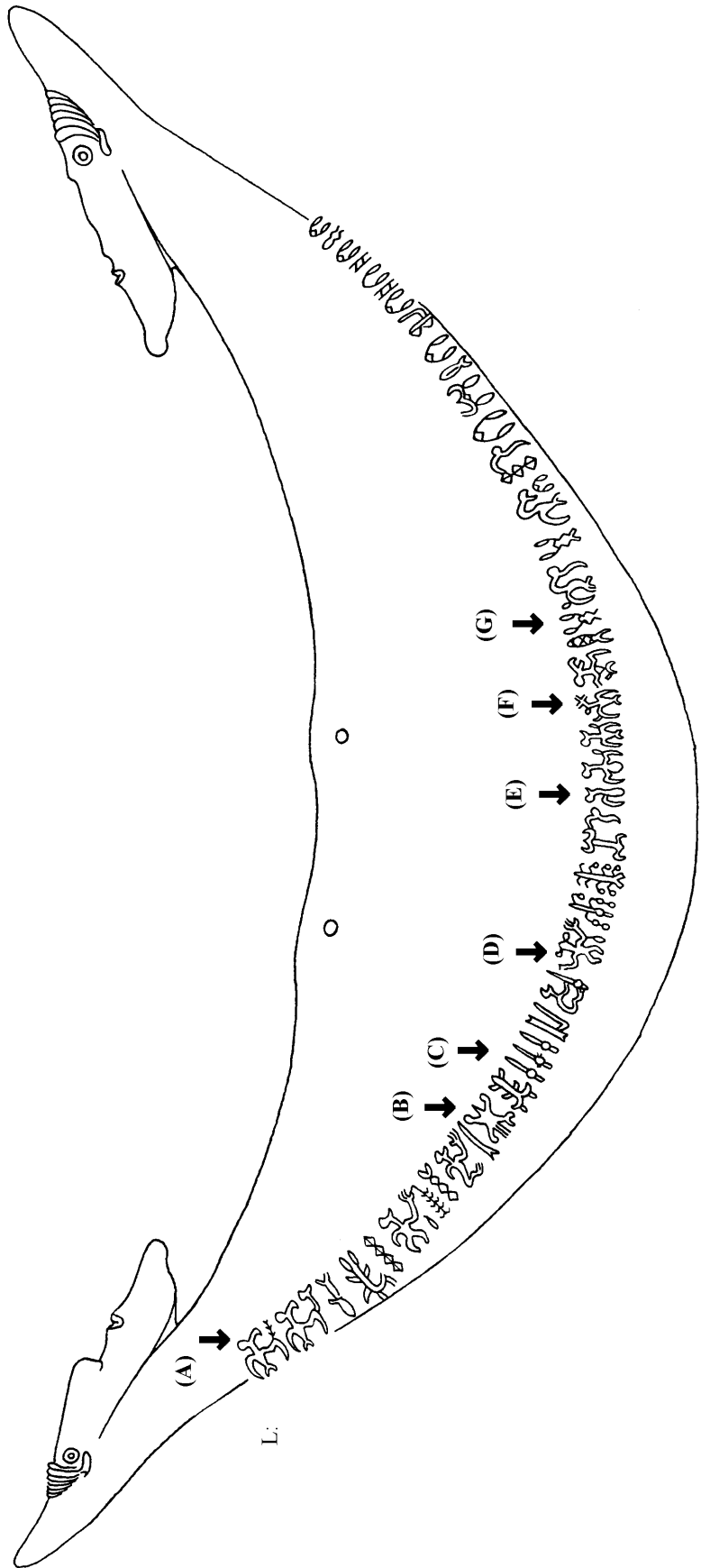

Figure 13.

# The Calendar and Mathematical Records in the *Rongorongo*

Consider the record on another London *rei-miro* (L), see figure 13. The text has been decoded (Rjabchikov 2009c). The following segments have engaged our attention: (A) **44 24 44 27 9** *tari Taha rau Niva* (the Frigate Bird, an image of sooty terns, is carrying the fruits, including eggs, from the legendary homeland Hiva), (B) **79** *Heke* (*Tuu-ma-Heke*, the legendary king and ancestor of the Miru tribe), (C) **135-135-135** *vera, vera, vera* (*vero* 'spear;' *vera* 'heat'), (D) **6-15** *Hora* (the month *Hora-nui*, September chiefly), (E) **68-68 44** *hohuihonui taha* (the great collection of birds = frigate birds literally), (F) **44-70 19 12** *tapu ki ika* (the prohibition of the fish), and a fertility formula: (G) **73 64 19 73 64 62 64 44b-17 64 23 64 12 64 18 4 64 4 64 4 64 73 64** *He mea ku, he mea toa, mea tua tea, mea ura, mea ika, mea tetu, mea atu, mea atu, mea, he mea.* 'I obtain, (I) obtain the sugar canes. (I) obtain the sweet potatoes (called) *tua tea*. (I) obtain the lobsters. (I) obtain the fishes. (I) obtain the big (fishes) (= the tuna fish etc.). (I) obtain the *atu* fish. (I) obtain the *atu* fish. (I) obtain. (I) obtain.).[2]

The arrival of sooty terns in the spring-time was a good sign that the inhabitants would receive copious eggs and fruits. It was the season of the increasing warm (cf. the words *vera, vera, vera*). According to local beliefs, the king of the island was a person who granted the abundance to all the people.

In the folklore text about the power of king the following mysterious formula is repeated many times: *Anirato maniroto* (Métraux 1937: 52-54). It can be read thus: *Anira to, manu (= mani) roto* 'Add (or take) now, that is inside the bird!'

Consider the record on the Aruku-Kurenga tablet, see figure 14.

Bv 8: 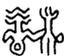

Figure 14.

Bv 8: **68 33 26-15** *Honu, vai* (or *ua*)*, Maro*. The Pleiades, the season of rains, the month *Maro* (June chiefly).

Here the word *Maro* (Old Rapanui *Maru*, *Marua*) is written as two syllables *ma* and *ro*. The New Year began with the new moon (June chiefly) after which the Pleiades rose first before dawn. The sequence of the Rapanui months gives *Maro* (June chiefly), *Anakena* (July chiefly), *Hora-iti* (August chiefly), *Hora-nui* (September chiefly), and so forth.

Consider the continuation of the list of the months on the same tablet, see figure 15.

Bv 10: 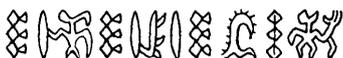

Figure 15.

Bv 10: **17 30-44 17 30-51-30 17 14 17 6-15**
*te Anakena, te Anakena, te Haua, te Hora*
(the month) *Anakena,* (the month) *Anakena,* (the moon) *Haua* (= the night *Atua*), (the month) *Hora*

The scribes avoided using signs for the sun (*ra*, *raa*) to write the syllable *ra*; they used the syllable *ro* (glyph **15**) because of very frequent alternations of the sounds *o/a* in the Rapanui language. The sign added to elbow of the raised arm of the figure of a man (glyph **6** *ho*, *ha*) in the last word does not read, this special mark denotes that the "arm" (glyph **15** *ro*) is a separate sign here. Glyphs **6-15** reads *Ho-ra*. (Glyph **6** represents a man, cf. Rapanui *hoa* 'friend;' but in many cases this sign reads *ha*.) It should be remembered that this tablet was a lesson book in the *rongorongo* school of king *Nga Ara*. The name *Anakena* (*Ana Kena*) is repeated twice in the text, and the word *kena* is rendered as the whole word (glyph **44** *taha*, *ta*; the designations of birds: *manu*, *kena*, *tavake* etc.) and as the pair of quasi-syllables *ke* and *(a)na*. All the names in the record are introduced by the definite particle *te*.

The name of the month *Maro* (*Maru*, *Marua*) could be written not only in the syllabic form, but also as an ideogram. Consider the record on the same tablet, see figure 16. In that text the winter solstice during the rainy season is described.

Bv 6: 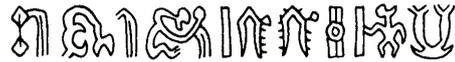

Figure 16.

Bv 6: **17 32 62 32 62 32 4 74-74 38 (= 34 39) 4-6 3 53** *Te ua, to(nga), ua, to(nga), ua. Atua tinitini raa tuha hina Maro*. Rains, rains, rains. (It was) the deity of the solstice during the time (*tuha*) of the month *Maro*.

The motif "two raised arms" is well known in the local rock art at the area called Mata Ngarau (The Painted Faces = *Makemake*'s images) in the religious centre of Orongo. The drawing of it and neighbouring symbols have arrested my attention (Lee 1992: 63, figure 4.31). First, the surrounded signs are bird-man figures and representations of birds (frigate birds). Second, the drawing of "two arms" is glyph **53** *Maro* (Old Rapanui *Maru*, *Marua*) carved on a rock. Third, beside this emblem the row of four cupules is incised. The months *Maro*, *Anakena*, *Hora-iti* (the small *Hora*) and *Hora-nui* (the big *Hora*) are four months from the beginning of the year in the local calendar. Thus, the month *Hora-nui*, when sooty terns arrived in the majority of cases (since the lunar calendar was used) and the bird-man was elected, was the fourth month. It is clear that the natives during the winter waited for the beginning of the *Hora-nui* month. The key indicator in the decoded rock figures is the record of the number four.

One can consider now the record on the same tablet, see figure 17.

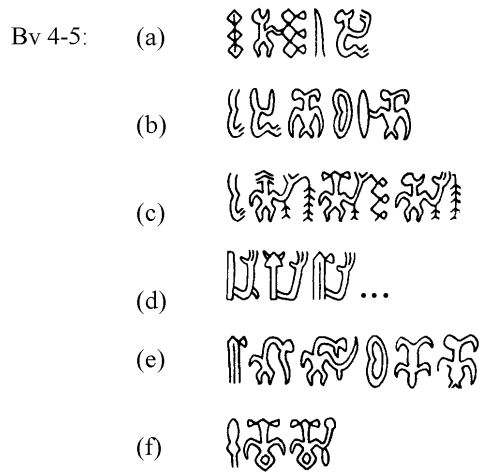

Figure 17.

Bv 4-5: (a) **17-4 17 5-6** *Te ha te tuha.* Four time intervals.
(b) **43 2 44 47 30-44** *Ma Hina Taha avae Anakena*. The moon goddess during the night *Tane* of the month *Anakena* went.
(c) **43 33 6-15 24 6-15 52 6-15 24** *Ma ua Hora-ari, Hora iti, Hora-ari*. The rains (together with) the bright sun of (the month) *Hora*, (otherwise) *Hora-iti*, (and with) the bright sun of (the month) *Hora* went.
(d) **4-15 21-15 26-15 …** *Atua roa Koro, Maro...* (They were) the great gods *Koro* (December chiefly; Father literally) (and) *Maro* (June chiefly)…
(e) **26 44 62 5 47 80 44** *Matatoa-atua ava, ui taha* (= *manu tara* figuratively). The warriors-lords elevated themselves, they watched the arrival of sooty terns.
(f) **73 68/65-68/65 4** *E Honu RANGI-Honu RANGI atua*. The great god: the Pleiades. (It was a celestial mark.)

I have decoded an instruction for teachers in the *rongorongo* school of king *Nga Ara* (Rjabchikov 2012a: 568-569, figure 8). Specifically, pupils wrote four lizard signs **69** *moko* together with the quasi-syllabic signs **26** *mo*, *ma*, *maa* and **21** *oko*, *ko*. The number of the lizards was brought out as the combination of glyphs **17-4** *te ha* (four). A variant of the article *te*, *ko te*, introduced the number one (*tahi*) in a Rapanui manuscript (Fedorova 1965: 398). In the read *rongorongo* text the beginning of the count was probably the new moon of the month *Maro*. *Koro* and *Maro* were days of the summer and winter solstices respectively. In the parallel fragment presented on the Great Washington tablet (Sa 5) both names of solstices are replaced with the term **26-4** *Matua* (Father) as a synonym for the term *Koro*. The warriors, the commanders of different troops, waited for the arrival of sooty terns. The Pleiades were the peculiar sign in the heavens during a long period.

Look at the sky above Easter Island again. In 1780 A.D. the Pleiades were seen almost precisely in the north before dawn on August 30. (August 20, 1780 A.D., beginning of dawn: 5:24 a.m., the Pleiades: azimuth: $08^0$ 51' 15"; August 29, 1780 A.D., beginning of dawn: 5:16 a.m., the Pleiades: azimuth: $00^0$ 45' 19"; August 30, 1780 A.D., beginning of dawn: 5:15 a.m., the Pleiades: azimuth: $359^0$ 52' 56"; August 31, 1780 A.D., beginning of dawn: 5:14 a.m., the Pleiades: azimuth: $359^0$ 00' 34".) In 1820 A.D. the Pleiades were seen in the north before dawn on September 1. (August 31, 1820 A.D., beginning of dawn: 5:14 a.m., the Pleiades: azimuth: $00^0$ 30' 53"; September 1, 1820 A.D., beginning of dawn: 5:13 a.m., the Pleiades: azimuth: $359^0$ 38' 40".) It was a signal of sooty terns at the beginning of September.

Consider the record on the Santiago staff, see figure 18.

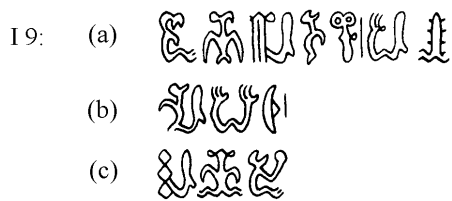

Figure 18.

I 9: (a) **62 44-26 (102) 62 60 (a vertical line) 58 (102) 25** *Too tama, too mata paoa tahi hua.* A young man took, a guard (warrior) took the first egg.
(b) **19 (102) 53 3 (a vertical line)** *Ki Maro hina* From the month *Maro*
(c) **17 (102) 69 4** *te Moko ha* (it was) the fourth new moon (*Moko = Hiro*, the first moon of each month).

Here the term *mata paoa* 'guard = soldier of a tribe' was used instead of the term *matatoa* (*mata toa*) 'warrior of a tribe.' In the text the basic topic of the bird-man festival is informed. A young man (servant) found and carried the egg of the sooty tern (*manu-tara*). This first egg was the incarnation of the god *Tiki-Makemake* in the local religious beliefs. The guard (warrior) who received that egg was declared the bird-man till the next elections.[3] We can see that the month *Hora-nui* (September chiefly) was fourth in the local calendar. In this text the word *mata* (tribe) is presented with the aid of the face glyph **60**, but in the record in figure 17 the same word is given as glyphs **26** *ma* and **44** *ta*. The vertical lines in the script were a late invention. They separated words to keep the text in the correct form. In the first case the line was written to read *tahi hua* (the first egg) merely, and the reading *mata paoa tahi* (the first guard) was unfeasible. In the second case the reading *hina (marama) tea* (the white moon) was not possible, too.

A *manu* (bird) song sounded on Easter Island in the old times:

*Aaku te ono i kouhau manu o te matangi o ure te ono* (Routledge 1914-1915).

It was the charm to gather (*ono*, *hono*) birds looking like a fertile (*ure*) wind. Rapanui *kohukohu* means 'storm, cloud,' *kohu raa* 'solar eclipse,' cf. the term *kouhau = kou hau* in the text, and the terms *hau* and *matangi* mean 'wind.'

I presume that such a chant must be inscribed on the staff. I have found it, see figure 19.

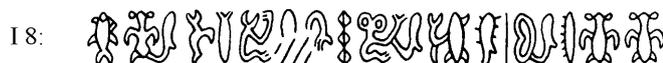

Figure 19.

I 8: **12var 68 (102) 72 9 6 48 15 17 60 (102) 123) 28 102 (a vertical line) 47 (102) 46 68-68**
*Aku (h)ono manu Niva hau, roa te matangi ure, ava na (h)ono(h)ono.*

Presently the birds from (the western land) Hiva are added (united, gathered) with the wind, the fertile wind grows, (it is) the ascent (*ava*) of the united (birds). {It is the new reading of mine on the base of new data.}

Glyph **12var** (**12**+**34**?) *ika* reads *aku* here (cf. the Rapanui names *aku*, *ihe aku*, *nanue para akuaku* of fish species). Old Rapanui *aku* 'presently' fits Mangarevan *akunei* 'ditto.' The term *matangi* (wind) was put down as the signs *mata-nga*. The term *hau* (wind) was put down with the help of syllables **6** *ha* and **48** *u*. Glyphs **102** and **123** are certain symbols of abundance and fertility, and they do not read in many cases. But here one glyph **102** is decorated with short lines as a badge and reads *ure*. Glyphs **68-68 44** *hohuihonui taha* rendered in figure 13 can just as readily be read *(h)ono(h)ono taha* 'ditto' due to the variations of the sounds. Thus, the song in Routledge's collection is a real quasi-bilingual text for the decipherment of Easter Island writing system.

Consider the record on the Aruku-Kurenga tablet, see figure 20.

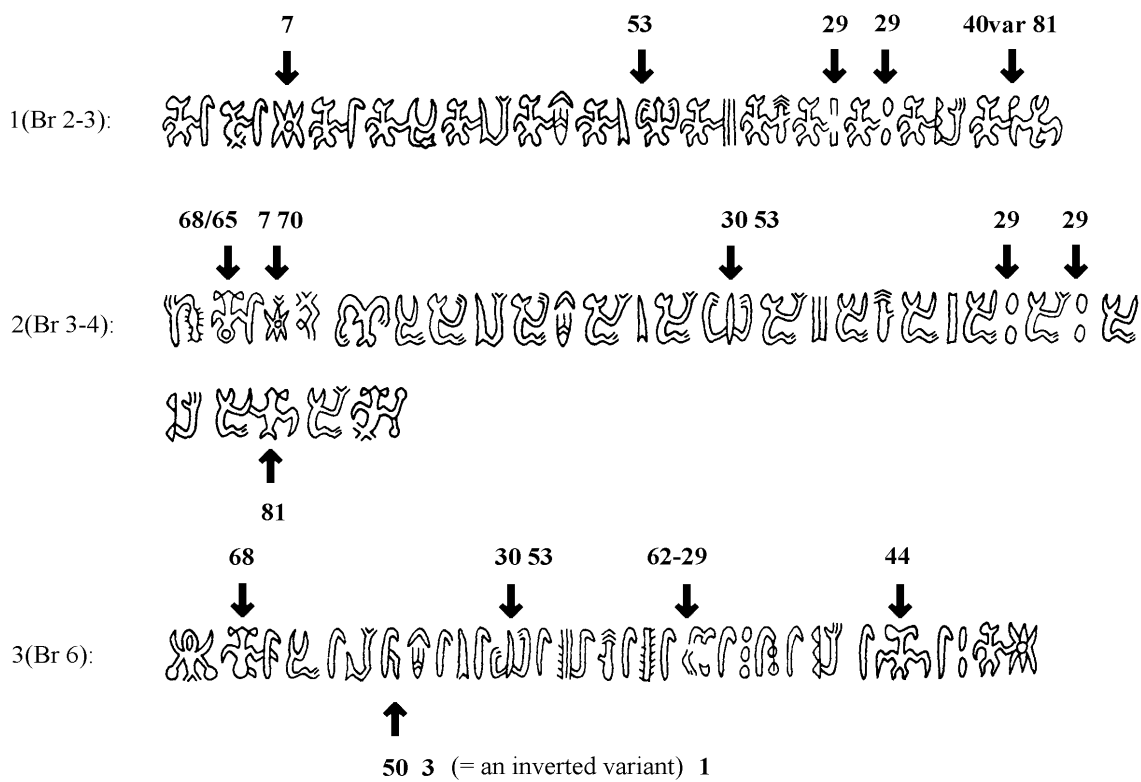

Figure 20.

I have analysed this record earlier (Rjabchikov 2016d: 2-3, figures 2-4). The motions of Aldebaran (**7** *Tuu*, **7 70** *Tuu Pu*) and the Pleiades (**68** *Honu*, *Hono*, *Ono*; where **65** *rangi* 'sky' is the generic determinative) in the month *Maro* (**53** *Maru*; June chiefly) are described. Since I held in my hands that splendid board, I can examine the text in detail. In fragment 3 glyphs **62-29** reads *toru* 'three.' The glyph looking like three rounds after the next sign **35** may be a variant of glyph **17** *tea*. On the other hand, it may be the determinative for the word *toru*. Consider fragment 1. Here glyphs **29** *rua* and **29** *rua* are inscribed. By the way, the second reading is questionable, it can be a version of glyph **17** *tea*. Be it as it may, at least the numbers two and three are found in these records. The number two gives such a result: it is the month *Anakena* (July chiefly), the next month after the month *Maro*. The number three gives the month *Hora-iti* (August chiefly). If both glyphs **29** *rua* and **29** *rua* are rendered in fragments 1 and 2 indeed, we shall receive the month *Hora-nui* (September chiefly). All three records terminate with the descriptions of arriving birds. They contain glyphs **40var 81** *(re)re manu* 'birds are flying,' **81** *manu* 'birds,' **44** *taha* 'frigate birds (sooty terns figuratively),' and the epithet referring to the sun god *Tiki-te-Hatu* (glyphs **6-4**, **6-7** *Hatu*, *Hotu*). It is the designation of the month *Hora-nui* (September chiefly). It is amply evident that the ancient priests-astronomers counted lunar months watching different phases of the moon as well as other celestial phenomena.

The onset of counting the time is the new moon of the month *Maro* in all the records. But it was an unusual year. Fragment 3 has the following segment: **50 3** (= an inverted variant) **1** *Hi* (= the 16th moon *Ina-Ira* as it is written in line Ca 7 of the Mamari record) *marama uri, Tiki*; it was the description of the partial (almost total) lunar eclipse before the sunrise on June 20, 1796 A.D.

The second *manu* (bird) song sounded on Easter Island in the old times:

*Matai epa kureri hoki te manu ture hau maru na te rangi na te werowero na te rere na te hohoku nui he atu hereri ai angaroi* (Routledge 1914-1915).

The corrected version of the song is as following:

*Mata(h)i e pa ku rere. Hoki te manu. Tuu, (re)re Hau(a). Maru na te rangi, na te verovero, na te rere, na te hohoku (= hohoki) nui. He atu(a) he rere ai (H)anga Roi.* The glyph (*pa*) of the first month (*ma-tai*, *ma-tahi*) (is here): the flight (of the birds) began. (The incantation:) Oh the birds, return! Oh (the moon goddess) *Haua*, come, fly! (It was the first month) *Maru* (June chiefly): (the incantation:) the sky (the first month), the spears (= heat) (the second month), the flight (the third month), the great return (the fourth month, September chiefly). The deities have been flown from the site (*ai*) *Hanga Roi* (*Hanga Oroi*? The bay of king *Oroi* from Hiva?).

The Rapanui folklore text "*Apai*" contains both terms *matai* (the first month) and *Maru(a)* (the first month, June chiefly) (Rjabchikov 1993a: 133-134). These terms were preserved in the Maori language: cf. *Maruaroa* 'name of the second month; and *Mātahi* 'first month; it begins with the first appearance of *Matariki* (Pleiades) before sunrise in June.' In the Rapanui calendar system those terms are united: *ku hakairi Maru matai, Maru matai.* 'The lunar month *Maru* (June chiefly) began (elevated itself literally).' The word *pa* (to write; to carve; glyph) is presented many times in a *rongorongo* lesson book known as the tablet Keiti (Rjabchikov 2013b). In the song the months from June till September are called with different epithets. It is apparent that the priest-astronomers stared at the starry heavens (*rangi*) to predict the beginning of New Year. The word *veravera* (heat) is presented in the *rei-miro* (L) record (see figure 13) as the thrice repeated word *vera* (heat). The arrival of sooty terns was a real signal of the growing warmth, many eggs, the bloom of plants, and the great harvest. The astronomical device at Orongo, the so-called "sun stones," served particularly to determine the day of the winter solstice exactly. According to Andersen (1969: 411), the Pleiades were connected with the spring and planting season in Polynesia.

Consider a segment of the archaic version of the famous song "*He timo te akoako*" on the Tablette échancrée (D), see figure 21.

Db 4: 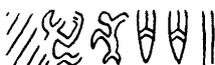

Figure 21.

Db 4 (Dr 4?): **6-19 1-1 4** *Hoki Tikitiki atua*. The great god *Tiki* returned.

In the version of the song "*He timo te akoako*" from the manuscripts that were published in the works of the Norwegian expedition (Heyerdahl and Ferdon 1965: figure 127) the last words *ara taha* (the road of frigate birds = sooty terns figuratively) correlate with the read record. One can cite a sentence from a Rapanui myth: *He hoki Makemake raua ko Haua.* '*Makemake* and *Haua* returned' (Barthel 1957: 72). *Makemake* was of an epithet of the sun god *Tiki* (*Tane*) (Métraux 1940: 314), and *Hau(a)* was an epithet of the moon goddess *Hina* (Rjabchikov 1987: 364-365, figure 2, fragment 5; 1988: 314-315, figure 1, fragments 5-8; 2014b: 165-166, 172, figure 9, fragment 3).

## New Readings of the *Rongorongo* Records Confirming My Decipherment

### (1) The Odd Form *ma-ha* Denoting Sharks

Consider the records on the Great St. Petersburg (P), Small Vienna (N) and Aruku-Kurenga tablets, see figure 22.

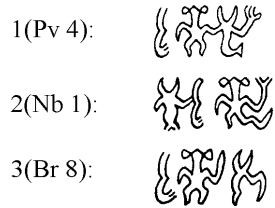

Figure 22.

1 (Pv 4): **43-6 11** *ma-ha*, or *ma-ho*, or *mo-ha*, or and *mo-ho* SHARK 'shark'
2 (Nb 1): **11 43-6** SHARK *ma-ha*, or *ma-ho*, or *mo-ha*, or and *mo-ho* 'shark'
3 (Br 8): **43-6 11var** *ma-ha*, or *ma-ho*, or *mo-ha*, or and *mo-ho* SHARK 'shark'

I have read earlier these texts as designations of sharks (Rjabchikov 1988: 318-319, figure 4, fragments 8, 9, 10), cf. Rapanui *mango* 'shark.' Glyphs **43-6** read not only *ma-ha*, but also *ma-ho*, or *mo-ha*, or *mo-ho*. The alternation of the sounds *h/ng* is possible in the Rapanui language. Glyph **11** reads *mango*, or *niuhi*, or *paki(a)*, or even *taoraha* etc. (shark, dolphin, seal, whale etc.).

On the wall of the famous cave Ana o Keke different signs are represented, and among them one can see the drawing of a whale or shark (cf. glyphs **11b**, **91**) and glyph **64** *mea* united together (Lee 1992: 47, figure 4.2; the interpretation in Rjabchikov 1994: 40, table 4), they read *ma-ha*, or *ma-ho*, or *mo-ha*, or and *mo-ho* (*mango*, *niuhi*, *paki*, *taoraha* and so forth) *mea* (red). Here the red colour was the emblem of the sea god *Tangaroa*.

In the Rapanui song "*Koro-rupa, te hare*" (Campbell 1999: 208-209) there are such words:

> *Ko te tai*:            The ocean (is round us):
> *Ie a hua, maho mea.*   (They are) solar rays (*i*, *hi*) of a calf (and)
>                         a whale. (The translation is of mine.)

I prefer to read the expression *maho mea* as 'whale' in this context because I have disclosed a solar symbol (cf. glyph **137a** *raa* depicting the sun with rays) on a drawing of a whale where three ones are shown (Lee 2004: 32, figure 2). To the third whale glyph **40** *(h)ere* is attached and it denotes that this mammal bears a calf. The solar symbolism of whales in the Proto-Polynesian times is discussed in Rjabchikov 2014b: 163.

Consider the record on the Santiago staff, where is rendered the description of elaborated carved drawings on slabs of the ceremonial platform Ahu Naunau at the royal residence Anakena, see figure 23.

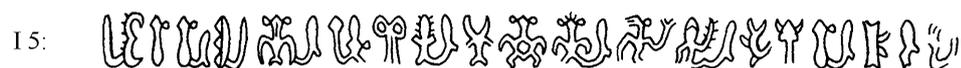

Figure 23.

I 5: **4-23 56-56 (102) 18-4 44-26 (102) 4 73 60 26 25 (102) 11 6-6 (102) 6 91 (102-123) 21 56 (102) 51 (102) 15**
*Tuura popo "Te atua Tama", "Atua e Mata" maa, "hua, Mango [Maho]", "Haha", "A Taoraha, oko", "Po kero".*
'The priests (*tuura*, *taura*, *tahura*) keep (the pictures) "The god *Tama* [*Tangaroa*]", "The god 'The Face'" made skilfully, "The whale-calf (and) the whale", "The darkness", "The whale (and) the whale-calf", "The dark night (The lizard)".' (Cf. Rjabchikov 2009a: figure 28, fragment 3. The repetitions of some motifs indicated their importance for king *Nga Ara*.)

One can attempt to find the form *maho* (*maha*, *moha*, *moho*) 'shark etc.' somewhere outside Easter Island. The personage *Ka moho-ali'i* (Shark god of the Pele family) [*Ta moho-ariki*] was mentioned many times in the Hawaiian folklore sources (Beckwith 1970: 90ff). The name has not been translated yet, although the meaning of the term *ali'i* (chief) is apparent. Standard Hawaiian *mano* (< *mago*) means 'shark.' But now it is clear that Hawaiian *moho* means 'shark' as well, cf. Rapanui *mango*, *mongo* 'ditto.' I think that the forms *maho*, *moho* and *mango*, *mongo* coexisted in the archaic times, and the forms *maho*, *moho* could be Proto-Tahitian variants (innovations).

Consider the record on the Berlin tablet (O), see figure 24.

O 4: 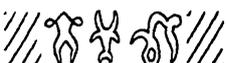

Figure 24.

O 4: (a damaged segment) **75 11 65/44b** (a damaged segment) … *ko Maho (= Mango) Rangi tua* … … the Shark 'Spica' of the sky was in the western sky …

As a parallel one can offer the words of the song "*He ika uru atua*" (Campbell 1999: 206-207) where in my view the total solar eclipse in September 16, 1773 A.D. is described in the religious terms:

| | |
|---|---|
| ... | … |
| *Ra te na.* | The sun is hidden. |
| *Ko He Ho, ko Maho-Rangi,* | The stars in the sky (are seen:) *He-Hoa* [the star *He* on the ecliptic] 'Castor or Pollux' (and) the Shark [*Maho = Mango*] 'Spica' of the sky. |
| *ko Rangi-Hetuu.* | |
| *Kohukohu Renga-Mitimiti.* | |
| *Ko te Nuahine Huri.* | (It is) the solar eclipse. |
| *Taua; a tae reka.* | (It is) the Black (*Huri = Uri*) Old Woman. |
| … | (It is) the egg (*toua*); (it is) not good. |
| | … (The translation is of mine.) |

Does the form *mango* (shark) sound exactly somewhere in the *rongorongo*? Yes, this record on the Santiago staff answers the question, see figure 25.

I 7: 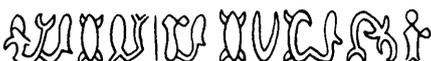

Figure 25.

I 7: **11 (102) 28 48-15 (a vertical line) 56 (102) 28 61 8 (102) 11 56/39**
*MANGO-ngo uri = Mango Uri. Po nga hina (= marama) Matua Mango po raa.*
(It was) the Black Shark (of the god *Tangaroa*) (= *Tapu mea*, the 24th night = *Tangaroa* with different epithets in some other Polynesian calendars). The night of many crescents (of the end of a lunar month) (called) *Matua* (= the 25th night) (has been produced by) the Shark (= *Tapu mea*, the 24th night) of the night/day.

I have demonstrated earlier (Rjabchikov 2009a: figure 14) that a syllable can be added to the ideogram to read the word correctly. In the Tahitian mythology the *ma'o uri* (black shark) is associated with the sea god *Ta'aroa* [*Tangaroa*] (Henry 1928: 355-359).

It should be pointed out that glyph **15** has the basic reading *ro*, but in some cases reads *ra* or *ri*, as such alternations of sounds were very frequent in the Old Rapanui language. For instance, Rapanui *aringa* 'face' has an unclear origin. But it reads *ari-nga* < *aro-nga*, where Rapanui *aro* means 'face; front.' Cf. also Rapanui *titi* 'to pile up' and *tito* 'frugal,' *tito koroiti* 'saving; economical;' *tingi* 'to beat; to hit,' *tingotingo* 'to beat to death,' *tingai* 'to kill' (<*tinga-i*); *veri* 'monstrous,' *veveri* 'execrable,' *vero* 'spear' as well as *karikau* (*kari kau*) 'hollow' and *karonga* (*karo-nga*) 'eye socket.'

**(2) *Te Haha*'s *Rongorongo* Records on the Paris Tablet (Snuffbox)**

Barthel (1958a) does not mention about this specimen of the script at all. A snuffbox made of pieces of a tablet (the Paris tablet) contains brief inscriptions in the classical style.

It is well known that Loti (1988) held a *rongorongo* tablet in his hands on the board French warship La Flore near the shore of Easter Island in 1872 A.D. I suggest that it was the Paris tablet.

One of panels (the bottom) of this specimen contains three brief lines of the *rongorongo*; I have featured the glyphs in figure 26.

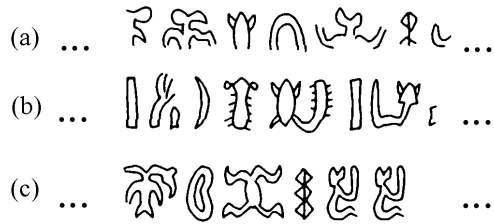

Figure 26.

(a) **… 60 44 28 39 (= 115, 139) 69 17 69 …** … *mata Taha Nga(a)ra Moko tea, Moko* … (The king of) the tribe of the Frigate Bird (the symbolism of the Miru tribe per Barthel 1978: 151) *Nga Ara* (of the tribal union) Moko (= Tuu, Momoko, Miru etc.) of the eaatern [and western] (parts of the island)…
(b) **… 4-14 3 12 56 28 25 4 5-21 4? …** … *atua roa Ika Hina ngau atua Tiko, atua*… the great goddess Fish-(the moon goddess) *Hina* bit the god *Tiki*…
(c) **… 44 47 108 17 6-6 …** … *Taha. Ava, hiri Te Haha*… ... The Frigate Bird. *Te Haha* elevates himself…

Let us look at the sky above Easter Island. One can calculate that a partial solar eclipse occurred on January 22, 1860 A.D. I suggest that this record says about the death of king *Nga Ara* soon before or after that event. It is common knowledge that that ruler died before the Peruvian slave raid of 1862 A.D. (Métraux 1940: 91).

*Te Haha* was a young man (boy) in the entourage of king *Nga Ara* (Routledge 1998: 242, 271; Rjabchikov 1999; 2012a: 567; 2012b). His name (**17 6-6**) is written down in segment (c). One can assume also that he (as a priest *ariki paka*) was a son of the king. After the death of his father, *Te Haha* could become a real ruler or co-ruler of Easter Island. Although the old man *Te Haha* aged more than 80 years told Routledge that he could not read and write the *rongorongo*, it was not the truth. Undoubtedly he was a *rongorongo* expert.

**(3) The *Rongorongo* Signs on Skulls**

At last a book has been published where symbols on skulls from Rapanui are seen well. Let us examine such sources of the script. The signs are presented in figure 27 (my own drawings).

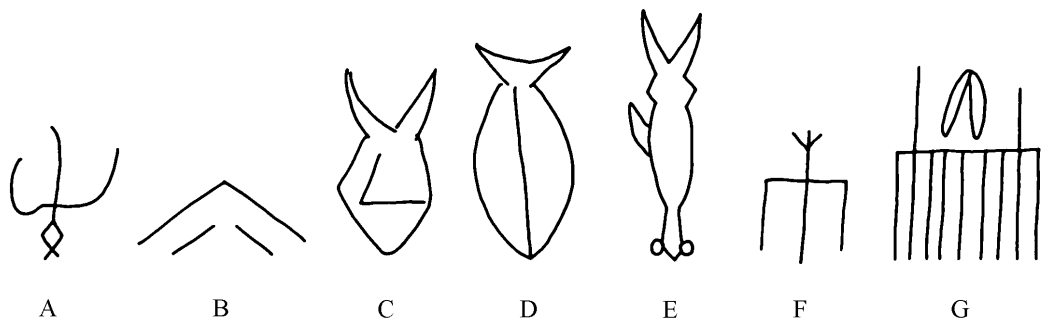

Figure 27.

1. Consider a complicated motif on the skull of 2-year-old child (Owsley et al. 2016: 260, figure 14.2), see figure 27A. I have read here glyphs **12 23** *ika Ura* 'the fish of (the deity) *Ura* (Lobster) = *Uraura-nga te Mahina* (an image of the dark phases of the moon).'

According to a Rapanui legend, a little girl once died; then she transformed herself into a fish; so, the girl cried that she in the hands of the spirits *Hiti Kapura* and *Urauranga te Mahina* (Blixen 1973: 10-11). A little boy was mentioned in the same situation in another narration (Métraux 1940: 372).

2. Consider a complicated motif on the skull of a female aged 35 to 39 years from Ahu Naunau (King's Platform) (Owsley et al. 2016: 262, figure 14.4), see figure 27B. I have read here glyph **149** *Hatuhatu* or *Hotuhotu*. It could be a late hoax: the name of the legendary king *Hotu Matua* was engraved on that skull to prove that the king had been buried there. On the other hand, this glyph could be a fertility symbol of a priestess or queen.

3. Consider a motif on the skull of a male (Owsley et al. 2016: 264, figure 14.6 [a]), see figure 27C. I have read here glyphs **16 149var** *Kahi Hotu* 'The Tuna Fish produces' or 'The Tuna Fish of *Hotu* or *Hatu* (= king *Hotu Matua*? or the god *Tiki-te-Hatu*?).'

4. Consider a motif on the skull of a male (Owsley et al. 2016: 264, figure 14.6 [b]), see figure 27D. I have read here glyph **16** *Kahi* 'The Tuna Fish.'

5. Consider a complicated motif on the skull of a male aged around 55 to 59 years that was disclosed under or near a ceremonial platform (Owsley et al. 2016: 264, figure 14.7), see figure 27E. I have read here glyphs **69** (the inverted head of the sign) **16** *Moko Kahi* 'the tribal union *Moko* (*Hanau Momoko*) of the Tuna Fish.'

In all three cases glyph **16** *kahi* (the inverted standard sign of the fish), prohibited as the meal for the majority of the people of Easter Island, was written. Various fishes represent the god of the sea *Tangaroa* in the Rapanui beliefs (Scheffrahn 1965: 58; Fedorova 1978a: 24; Rjabchikov 2014b: 172). This god was also known everywhere in Polynesia as *Tama* (*Tanga*) 'Father; Ancestor' with different epithets (Rjabchikov 1988: 316-317, figure 3, fragment 9; 2014a; 2014b: 172-173, figure 9, fragment 6; 2016b; 2016c). According to Métraux (1940: 127), that god (*ariki*, king) was the forefather of the Rapanui kings from the Miru group. I conclude that all three records were made on the skulls of kings of the Miru group (Tuu, Moko, Hanau Momoko). These crania had the great *mana* (the supernatural power) per local beliefs. In compliance with the Rapanui legend "*Puoko o te ariki*" (Englert 2002: 144-145), 'the sign of a tuna fish as the sign of an inverted fish' (*hai kahi, hai ika hoki*) was incised on a king's skull.[4]

Consider in this connection the record on the Small Santiago tablet, see figure 28.

Gr 1-2: 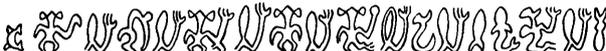

Figure 28.

**Gr 1-2: 8 6 35 48-15 44b 48-15 6-15 48-15 6 48-15 6 48-15 27 48-15 12 11 72 48-15 16**
*Matua a* ADZE (= *Tara-i* or *Tara*) *uri tua, uri Tangata roa* [= *Tangaroa*], *uri Tangata, uri Tangata, uri Rau, uri Ika Pakia Manu, uri Kahi* (= *Tangaroa*).
*Matua* (son of) *Tara*, son of the open sea, son of *Tangaroa* [The great man, the great ancestor], son of a man, son of a man, son of *Tinirau*, son of *Rongo-ma-Tane*, son of *Tangaroa*.

Here a genealogy of the explorer-king *Hotu Matua*, son of explorer-king *Tara tahi* (The first *Tara*), who both arrived from Mangareva (Rjabchikov 2014c; 2014d). His family line went back to the god *Tangaroa*. The memory about *Tara tahi* was retained in the image of king *Tuu-ko-Iho* (*Tuu-ko-Ihu*). In the Rapanui legend "*Te moai kavakava a Tu'u-ko-iho*" (Englert 2002: 102-107), the terms *kautoki* (*kau-toki*) and *tarai* (*tara-i*) mean 'adze' and 'to carve' respectively.

6. Consider a complicated motif on the skull of a female aged 20 to 24 years (Owsley et al. 2016: 266, figure 14.9[a]), see figure 27F. Glyphs were incised on this skull when its face was turned from the scribe: they are glyphs **26-15** in the cursive style reading *Maro* and denoting in this case the islet Marotiri. During the horrible wars between the tribal unions Tuu (Miru etc.; Hanau Momoko) and Hotu-iti (Tupa-hotu etc.; Hanau Eepe) some beautiful young women of the Tupa-hotu tribe who hid themselves on the islet Marotiri were killed (Métraux 1940: 74-84). Perhaps the skull of such a woman was found many years later; on the cranium the name of the islet was engraved.

7. Consider a complicated motif on the skull of a male aged around 40 to 44 years (Owsley et al. 2016: 266, figure 14.9[b]), see figure 27G. A drawing (glyph?) of a grass skirt (cf. Rapanui *kahu* 'dress') and glyph **64** *mea* are engraved here. I presume that this record reads *Kahu Mea*.

According a legend (Englert 1974: 99-100), *Kahu Mea* was a Tupa-hotu nobleman, son-in-law of the Tupa-hotu chief (king?) *Kauaha*.

I have read earlier a peculiar glyph consisting of four parallel lines as a variant of glyph **34** *raa* (the sun) (Rjabchikov 2009a: figure 40). Obviously it was a different glyph with the reading *kahu* (now it is glyph **174**). Consider the record on the Santiago staff, see figure 29.

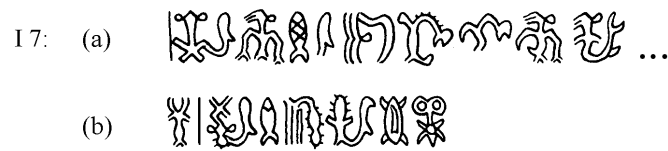

Figure 29.

I 7: (a) **(a vertical line) 49 (102) 6 12 (102) 33-29 62-62 6 11 …** *(Ariki) mau a ika vairua Toto a Mango…*The killed king as the ghost *Toto a Mango…*
(b) **11 (a vertical line) 70 56 (102) 12 174 25-25 5/28 60/7 …** *Mango pua Poike Kahu. Huahua tinga mata Tuu … Mango (Toto)* killed *Kahu (Mea)* from *Poike*; (then) the sons (of the latter character) killed (the people) of the tribe *Tuu*… {This segment has been decoded anew on the basis of the new data.}

According to this record, the hero (king) *Mango Toto* of the western tribes killed *Kahu Mea* during the terrible war around 1682 A.D. The name *Mango Toto* means 'The Bloody Shark.' It retained in the name of the ghost *Nuku te Mango* (The army of *Mango* [*Toto*]). Forster (1777: 591) wrote the name of the statue Mangototo [*Mango Toto*]. In accordance with Rapanui legends, ten years the authority belonged to the Miru, then about ten years the authority belonged to the Tupa-hotu (Métraux 1940: 382).

8. Consider four circles that were cut from the skull of a female aged 20 to 24 years perhaps from the ceremonial platform Ahu Naunau (Owsley et al. 2016: 259, figure 14.1). I surmise that such rounds had the symbolic value and represent glyphs **115** (**139**, **39**) *taka*, *raa*. It was probably the skull of either a priestess of the solar cult or a princess (queen). Per Thomson (1891: 497), at Anakena (in the other words, in the vicinity of the ceremonial platform Ahu Naunau) a female statue called *Viriviri Moai a Taka* was situated. Barthel (1958b: 254) did not find such a monument there. So, I suggest that *Viriviri a Taka* (The Motion of the sun) was the name of that priestess or princess (queen) in fact.

### (4) The "Sun Stones" are Indirectly Described on the Tahua Tablet

The "sun stones" were a local device at the ceremonial centre at Orongo to determine days of solstices and equinoxes (Ferdon 1961; 1988). A report about such measurements is preserved in the record on the Tahua tablet (cf. also figure 17), see figure 30. I have corrected the drawing.

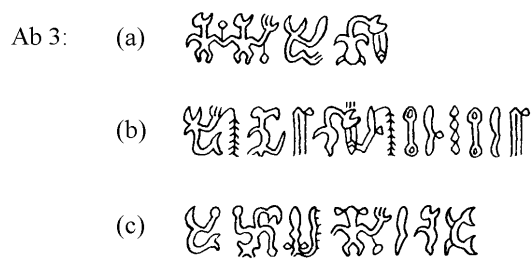

Figure 30.

Ab 3: (a) **6-40-6-40 6 68/62 1** *Harehare a Honu, too Tiki*. (It was) the big House of the Pleiades; the (sun god) *Tiki* took (it).
(b) **11-24 19 26 62 1 50 24 65 50 17 65 50 26** *Maho (Mango)-ari, ku maa: too Tiki hi ari RANGI, hi tea RANGI, hi maa*. (It was) Spica (associated with the sun god *Tangaroa*); the brightness of the sun increased: (the sun god) *Tiki* took the clear rays THE SKY, the clean (white) rays THE SKY, the bright rays.
(c) **2 62 73 50 6-15 50 19 8 …** *Hina too, hei Hora. Hiki Matua…* The moon (the moon goddess *Hina*) took, (this goddess) drove the month *Hora* (*Hora iti* or *Hora nui*, August or September for the major part). The Canoe (β and α Centauri) was rising (in September)…

Here the features of the months (a) *Maro* (June chiefly; the day of the winter solstice), (b) *Koro* (December chiefly; the day of the summer solstice) and (c) *Hora-nui* (September chiefly; the day of the vernal equinox) are mentioned.

The read text was a fragment of an exercise at a late stage in education in the *rongorongo* school of king *Kai Makoi* the First. The pupils repeated glyph **50** *(h)i* several times. On that lesson glyphs **11 24** *Maho (=Mango) ari* 'Spica' were basic.

The similar exercises on those glyphs in the same school are presented on the Keiti tablet (Rjabchikov 2013b: 9, figure 6; 11, figure 10), see figure 31, fragments 1 and 2. In fragment 3 the record on the Tahua tablet is presented that precedes the record in figure 30. During a lesson the students wrote the name of Spica many times.

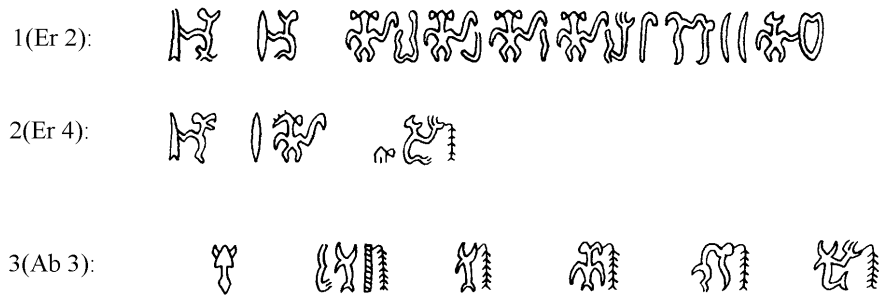

Figure 31.

1 (Er 2): **5-19 30 62 6 35 50-50 6 35 50 6 35 50 6 35 50-15 35 50 44b 3 3 6 47** *Tuki ana, to: ha pa: **hihi**, ha pa: **(h)i**, ha pa: **(h)i**, ha pa: **ira**, pa: **i tua**, marama, marama, a avae*. Carve (= write) a lot, add: you have carved: (glyphs) **50-50** *hihi*, you have carved: (glyph) **50** *(h)i*, you have carved: (glyph) **50** *(h)i*, you have carved: (glyphs) **50-15** *ira*, carve: (glyphs) **50 44b** *i tua*, (carve) during a day, (carve during another) day, during (this) month.
2 (Er 4): **5-19 30 49 35 26-6 24 …** *Tuki ana:* (Carve = write a lot:) *mau pa* (the abundance [*mau*] of carvings = signs [*pa*]) (the words:) **26-6 24** *Maho ari* (Spica)…
3 (Ab 3): **21 43 11 4 24 11 24 44 24 44b 24 11 24** *Ako ma-MAHO atua ari, Maho ari, taha: ari, tua: ari, Maho ari*. Learn (the glyphs) **43 11 4 24** *Maho atua* (deity) *ari*, (the glyphs) **11 24** *Maho ari*, turn (a tablet, a banana leaf): (the glyphs) **24** *ari*, (write) on the other side (of the tablet or that leaf): (the glyphs) **24** *ari*, (the glyphs) **11 24** *Maho ari*.

**(5) The Farther Analisis of Reports about Elections of Bird-Men**

I have deciphered the record on the Great St. Petersburg tablet (Rjabchikov 2013a: 6, figure 7). This text and its parallel record on the Small St. Petersburg tablet (Q) are presented in figure 32.

1 (Pr 6): **105 68 17 3 1 6-4 19 (102) 6-6 56 6-6 77 62 15 115-115 6 12 44-44 49 21 49 35 25 6 3 49 4/33 6 (123) 6-51 44-33 48 7 102 6 11 6/26** *Moe, hono te Hina, Tiki, hatu. Kio haha, paoa haha. Mama too ro takataka. (H)a ika tahataha mau oko, mau pa hua (h)a Hina. Mau atua/ua (h)a ake. Tau Utu ure a Mango-Ama*. The moon goddess (and the sun god) *Tiki* slept (and) were united, they produced (the eggs). The servant (*mata-kio*) took (the egg), the warrior (*paoa, mata-toa*) took (the egg). (The deity by the name of) *Maamaa* ('The bright colour') took the red hot colour (or the sun figuratively). The bird-man ('corpse-frigate bird') held the egg ('ripe fruit'), (he) held the egg (*hua*) laying (on a piece of tapa) of the moon goddess. The deity (= the bird-man) held (a symbol) of abundance. (It was) the year of *Utu*, son of *Mango Ama* (or *MAHO Maho*?).

2 (Qr 6-7): **105 68 17 3 1 6-4 19 (102) 6-6** (a damaged segment) **6 35 56 6 3 49 59/33 6 6-51 44-33 48 7 102 6 11 6/26** *Moe, hono te Hina, Tiki, hatu. Kio haha,* (a damaged segment) *ha pa po (h)a Hina. Mau kaua (h)a ake. Tau Utu ure a Mango-Ama.* The moon goddess (and the sun god) *Tiki* slept (and) were united, they produced (the eggs). The servant (*mata-kio*) took (the egg), (a damaged segment) (the bird-man) touched the round object (*popo* = the egg) laying (on a piece of tapa) of the moon goddess. The progenitor (= the bird-man) held (a symbol) of abundance. (It was) the year of *Utu*, son of *Mango Ama* (or *MAHO Maho*?).

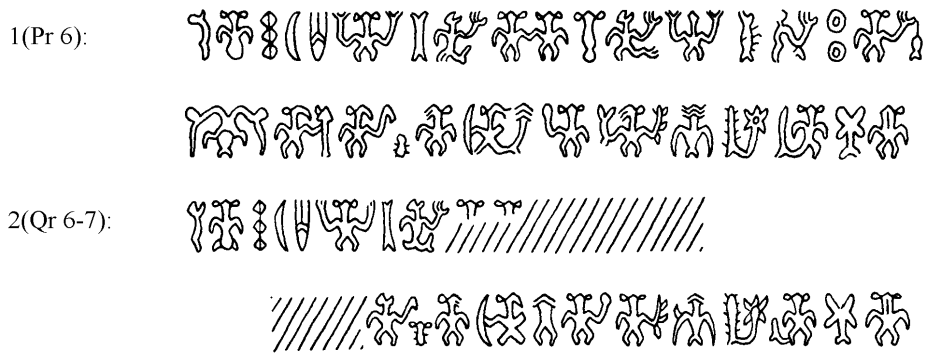

Figure 32.

The verb *mau* 'to hold' (glyph **49** *mau*) in fragment 1 is replaced by the verb *ha* 'to touch' (glyph **6** *ha*, cf. Rapanui *haha* 'to touch tentatively').

The key indicator of these records is the word *ake* 'abundance' (glyphs **6-51**), it is presented in figure 33. Call attention to the two versions of glyph **6** *(h)a*.

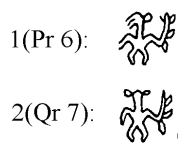

Figure 33.

Old Rapanui *ake* means 'more' (Fedorova 1988: 107), cf. Tuamotuan *ake* 'ditto.' In the texts of two Rapanui songs in the common environs the place names *Toka-Miti-Ake* and *Toka-miti-miti* are mentioned (Campbell 1971: 400-401, 453). In the Rapanui language the reduplications are often used to augment the force of the morpheme (Fedorova 1978b: 47). Hence, *miti-ake* = *mitimiti*, and Old Rapanui *ake* signifies 'more; abundant; abundance; many etc.' Both place names mean 'Underwater rock in the salt water.'

I agree in general with Horley's (2012: 58-61, table 1) list of the names of bird-men (97 entries) composed on the base of Routledge's lists. The bird-man called *Utu* in the records in figure 32 was known as *Utu-Piro*; he ruled in ca. 1850 A.D. (Métraux 1940: 339). If we assume that this dating has been estimated exactly, the last bird-man *Rukunga* or *Rokunga* ruled in 1867 A.D. Hence, the bird-man called *Ko Hake* (= *Ko Ake* 'The Abundance') ruled in 1771 A.D.

The total solar eclipse occurred during the bird-man festival on September 16, A.D. 1773. The name of the bird-man was *Hopu Eroero*, where the last word is comparable with Rapanui *eoeo* 'ashes' (the alternations of the sounds r/- were possible). Maybe the situation was so bad (due to the disappearance of the sun during the feast) that a servant (*hopu*) was declared as this bird-man. Thus, he was named after the eclipsed sun.

The simple calculation shows that the bird-man *Kohu te Tangi* from the Miru tribe ruled in 1860 A.D. The partial solar eclipse occurred on January 22, 1860 A.D. The name means The 'Eclipse Cries' (perhaps because of the death of king *Nga Ara*; see above), cf. Rapanui *kohu raa* 'solar eclipse.'

The partial solar eclipse occurred on December 31, 1842 A.D. The bird-man *Ko te Piko Kohu* (Hidden, Eclipsed) could rule in 1843 A.D. His number in Horley's list allows calculating the date of 1842 A.D.

The partial solar eclipse occurred on September 7, 1839 A.D. The bird-man *Ko Hiti ko te Vai Kino* (The Bad Water Rose; the symbolism of the darkness and rains) ruled that year.

The partial (almost total) solar eclipse occurred on March 25, 1838 A.D. The bird-man *Kohe Po ko Rano Kao* (The Darkness [*Koe*, *Kore*] of the Night at Rano Kao/Kau) ruled in 1838 A.D.; that eclipse had been predicted. (It is my hypothesis yet.)

The partial (almost total) solar eclipse occurred on October 29, 1818 A.D. The bird-man *Ko Raua Kere* (That who Became [was Created] Black) ruled in 1819 A.D. His number in Horley's list allows calculating the date of 1818 A.D.

The partial (almost total) solar eclipse occurred on August 5, 1804 A.D. The bird-man *Ko Ta-Uru Ehu* (The Ashes or Smoke Entered) ruled that year.

So, a number of bird-men were named after the eclipsed sun.

The presence of several bird-men of the eastern tribes from 1803 till 1808 or 1809 A.D. could imply that the all-powerful king *Kai-Makoi* the First (*Ko To Hi* [As the Sunbeams] *Tai/Kai*) died at that time. The authority of his son *Nga Ara* was unstable.

According to Routledge (1998: 265), the last bird-man *Rokunga* ruled in 1866 or 1867 A.D.

These data show that we have obtained the quite precise chronology (the error was about one year) of Rapanui ancient history from 1771 till 1867 A.D.

### (6) How I Have Suddenly Realised a *Rongorongo* Text

Only now I understand the record (see figure 34) on the Tahua tablet, though I read it many years ago.

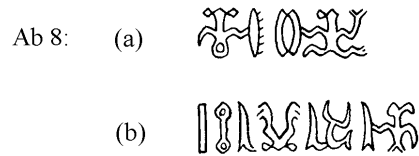

Figure 34.

Ab 8: (a) **68/65 46 57 69** *Honu RANGI naa tara, moko*. The Pleiades are (still) invisible before dawn (in June).
(b) **4 65 5 53var 5 2 5 44** *Atua Rangi, atua Maru (Maro), atua Hina, atua taha (= manu)*. The deities (are) the Sky (the description of the month *Maru* according to the second *manu* (bird) song, see above), the month *Maru* (*Maro*) [June chiefly], the Moon goddess, the birds (sooty terns).

It was an incantation read during the cold season. The natives waited for the arrival of sooty terns which were a certain symbol of the growing warmth of the sun.

### Conclusions

There is a good cause to assert that the Easter Islanders constantly watched Aldebaran and the Pleiades in the past. The Russian scholar Irina K. Fedorova and the American scholar Georgia Lee were the first who contributed significantly to the archaeoastronomical studies of these celestial bodies. I have decoded the Mataveri calendar completely. The ceremonial platform Hekii 2 was oriented on Aldebaran and nearby stars. The disappearance of the Pleiades during the dawn period in the north at the end of August could be an important mark before the arrival of sooty terns and before the elections of the next bird-man. The calculated dates of several solar eclipses have been used for composing the chronology of Easter Island from 1771 till 1867 A.D. The ancient priests-astronomers watched the sun, the moon, Spica as well as β and α Centauri as well. At least two inscriptions, on the Tahua and Aruku-Kurenga tablets, indirectly witness that the "sun stones" were a special astronomical device at the ceremonial centre of Orongo to determine the days of solstices and equinoxes.

## Acknowledgements

I wish to thank Fr. Paul Lejeune for his kind permission to study four excellent *rongorongo* tablets (Mamari, Tahua, Aruku-Kurenga, and Tablette échancrée) in the General Archives of the Congregation of the Sacred Hearts of Jesus and Mary (Rome) in May 2015. I am grateful to Ms. Luana Tarsi for her assistance during that research.

## Notes

1. Old Rapanui *toke* (= *toko*) 'great god; deity' corresponds to Maori *toko* 'sacred pole or stick set up in honour of a deity.' Moreover, Maori term *toko* denotes several gods including *Tane* and *Tangaroa* (Te Rangi Hiroa 1949: 467). *Takapau* was the name of that statue. *Taka* means 'round = the sun' and *pau* (*pahu*) means 'signs, carvings.' So, it was really an image of the sun god (*Rarai-a-Hova*, *Rarai-a-Hoa*, *Makemake*, *Tiki*). Another name of this house was *Taura-renga* (Métraux 1940: 106). I suggest that this expression, *Taura Renga*, means 'The priests *taura* of the sun (*Renga*, the Yellow Colour).' The name of the same statue called *Hoa-hakananaia* means 'The Friend (an epithet of the sun deity) moves quickly.'
2. About fish *atu* (Old Rapanui *atu* ['*atu*] 'bonito or skipjack tuna; *Katsuwonus pelamis*) see Rjabchikov 2011b. Perhaps the phrase *tapu ki ika* in this record describes primarily the tuna (*kahi*) fish.
3. Glyph **56** *po* represents the club *paoa* (Rjabchikov 1987: 366, appendix) and reads *pao(a)* and even *pa* in some records. Routledge (1914-1915) recorded a number of names of bird-men. Here I offer the translation of several such names: *Ko Paoa ko te Manu-Tara e Tahi* (The Guard of the First Sooty Tern = The Guard of the First Egg), *Pua Ua* (The Egg of the Dwelling), *Ko Pua Hau* (The Egg of the King; or the Egg of the Moon Goddess), *Ko Pue ko te Manu* (The Egg of the Bird = the Egg of the Sooty Tern), *Hihia Pua Moko* (The Lift of the Egg of the tribal union *Momoko*), *Ko Tahi o Hiva* (The First from Hiva), *Ko te Ara Hiva* (The Road from Hiva), *Tahi a Marua* (The First Month *Marua* = *Maru*), *Manu Api* (The Hidden Bird), *Ko te Manu Renga* (The Bird of the Sun), *Ko Taha Oi* (The Motions of the Frigate Bird in Different Directions), *Tahi Ao Ria* (The First Strong Authority or The First Strong Ceremonial Paddle; *ria* = *riri*), *Ko Hina a Vaivai Tea* (The Moon Goddess as a Source of the Clean Water), *Ko Hina Mango* (The Moon Goddess of the Shark; cf. the same name of a priest connected with king *Kai-Makoi* the First), *Koro Henga* (The Month *Koro* of the Bright Sun), *Ta-kero* (The Darkness), *Ko Ke Hunga* (The Hidden [Person] or the Disappearance), *Tuu Hotu Roa* (The Great Abundance Came), *Mata Popo Ra* (The Face of the Hot Sun, cf. Rapanui *popo* 'round'), *Mata Rorerore* (= *Mata Roa*, *(Re)Re*, The Growing (and) Flying Face = the sun, the arriving sooty terns), *Ngata Hora* (The Man of the month *Hora*), *Ko te Mau Taka-taka* (The Abundance of the Sun), *Ko Niu Ka-ka Vera* (The Nut [Egg] Associated with the Growing Warm), *Ko Hihi* (The Solar Rays), *Taku Riko* (The Count or the Prediction of *Riko*), *Hengahenha ma Taka* (The Bright Light for the Sun), and *Papa Haha Raa* (A Rock for Watching the Sun).
4. It is the translation of mine. The literal meaning of the expression *ika hoki* is 'the fish returning (= diving into the sea).' Hence, in the reality the tuna fish were available for the local aristocracy only in the past.

# Appendix

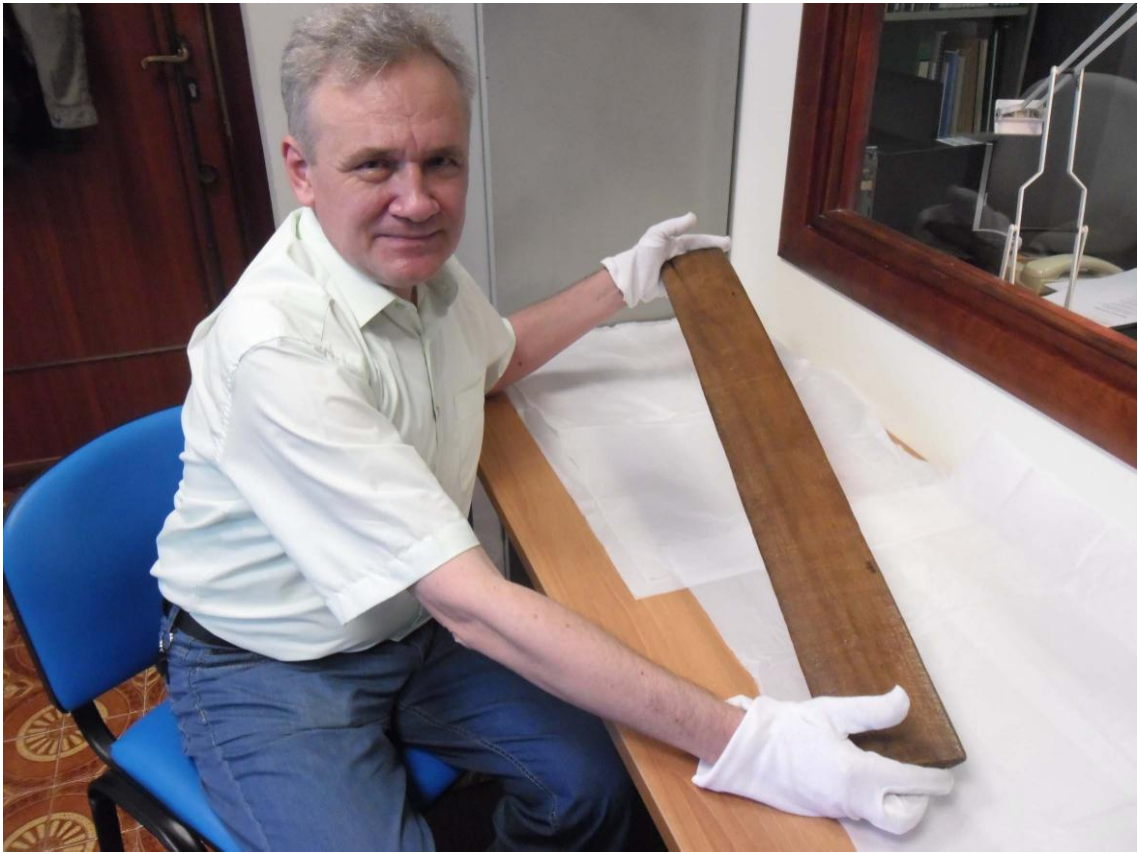

I visited the General Archives of the Congregation of
the Sacred Hearts of Jesus and Mary in 2015.
I was tracing the *rongorongo* glyphs of the Tahua tablet.